\documentclass[prl,aps,twocolumn,epsfig,floatfix]{revtex4}

\usepackage{epsfig}

\newcommand{\xx}{{\bf x}}

\begin{document}

\title{Correlations in atomic systems: Diagnosing coherent superpositions}

\author{Radka Bach}
\author{Kazimierz Rz{\c{a}}{\.z}ewski}
\altaffiliation[Also at]{
	 Faculty of Mathematics and Natural Sciences, Cardinal Stefan Wyszy{\'n}ski University, Warsaw, Poland}
 \affiliation{Center for Theoretical Physics, Polish Academy of Sciences, al. Lotnik{\'o}w 32/46,
02-668 Warsaw, Poland}

\date{\today}

\begin{abstract}
While investigating quantum correlations in atomic systems, we note that single measurements contain
information about these correlations.
Using a simple model of measurement --- analogous to the one used
in quantum optics --- we show how to extract higher order correlation functions from individual
"photographs" of the atomic sample. As a possible application we apply the method to detect a subtle phase
coherence in mesoscopic superpositions. 
\end{abstract}

\maketitle

Correlation functions provide a useful tool for investigating complicated states, both classical and quantum. 
For example, measuring intensity-intensity correlations for photons has been a significant development in
astrophysics \cite{Brown&Twiss} and investigation of the second order correlation function has led to understanding 
of antibunching of photons \cite{antibunching}.
For atoms, correlation functions might prove to be as useful as they have been for photons. There are
certain important differences, however. 
For light, if the underlying quantum state is unknown, 
the knowledge of a correlation function of a given order does not determine lower order
correlation functions and a separate experiment must be
set up to measure the latter \cite{Walmsley1,Walmsley2}. This is not true in atomic case. For atoms a
superselection rule holds, which prohibits the atomic state to exist in a superposition of Fock states with
different number of atoms. This proves to be a very strong condition and implies, as will be shown in more
detail, that there exists a unique hierarchy of correlation functions. An important consequence is 
that it is possible to extract information about higher order correlation functions, not only density, from
typical measurements performed in atom optics.
In this Letter we derive this hierarchy of correlation functions and provide a simple
prescription for extracting correlation functions of interest from typical measurements, which we model in
an analogous way to quantum optics.
As an illustration
we use the method as a diagnostic tool to distuinguish a coherent superposition against a simple mixture.

There exists a number of possible detection schemes in atom optics -- via resonance fluorescence, 
non-resonant imaging or ionizing the atoms and detecting the resulting electron current, just to name a few -- 
and each of them requires a slightly different theoretical approach \cite{Meystre}. In particular one cannot
claim that all measurements of atomic samples correspond to normarly ordered correlation functions 
and in that respect atom optics differs from its photonic
counterpart, in which all correlation functions are normally ordered since light
is detected via absorption. On the other hand, however, the notion of coherence, that has been originally introduced
for photons in the context of normally ordered correlation functions and their factorization \cite{Glauber}, 
is a very useful tool and it would be desirable to classify atomic systems in an analogous manner.
Therefore we will further assume that the measurement of the atomic cloud is done in such a way that 
normally ordered correlation functions are relevant, even though it might be an idealization of an actual
detection scheme.

Consider a state composed of $\cal{N}$ atoms and described by a ket $|\psi\rangle$ for 
a system with a second-quantized field $\hat{\Psi}(\xx)$. 
The diagonal part of the $r$-th order correlation function is then defined as:
\begin{eqnarray}
& & 
	G^{(r)}_{|\psi\rangle} (\xx_1,\ldots,\xx_r) := 
\nonumber \\ & & 
	\left\langle \psi \left|  
	\hat{\Psi}^\dagger(\xx_1) \cdots \hat{\Psi}^\dagger(\xx_r)
	\hat{\Psi}(\xx_r) \cdots  \hat{\Psi}(\xx_1) 
	\right| \psi \right\rangle
\end{eqnarray}
and, within the assumptions described above, is proportional to the joint probability of detecting $r$ atoms: 
one at position $\xx_1$, one at $\xx_2$ and so on (of course the positions need not be all different). 
Time dependence has been dropped for simplicity: we assume that all atoms are 
measured at the same instant. In other words we neglect the state's dynamics during the time of measurement.

Provided the system is in a Fock state, i.e.
$|\psi\rangle$ is an eigenstate of the total number of atoms operator 
$\hat{\cal{N}} = \int d\xx \; \hat{\Psi}^\dagger(\xx) \hat{\Psi}(\xx)$,
a simple algebra shows that the following relation holds both for bosonic and fermionic systems:
\begin{eqnarray}
& &
	\int d\xx_{r} \; G^{(r)}_{|\psi\rangle} (\xx_1,\ldots,\xx_r) = 
\nonumber \\ & &
	\left( {\cal{N}} - r + 1 \right) 
	G^{(r-1)}_{|\psi\rangle} (\xx_1,\ldots,\xx_{r-1}) 
	\label{eq:hierarchy_small}
\end{eqnarray}
Thus, a correlation function of a given order contains all lower order correlation functions.
Such a hierarchy exists for any given state, but only for Fock state is the coefficient 
of proportionality so simple and general.
Assumption that the state is in a Fock state is valid 
for most experiments performed with cold atoms, Bose-Einstein condensates in particular, since
they are almost perfectly isolated from the environment. Moreover, if one was to allow for interactions
with the particle reservoir, the resulting state is a mixture of Fock states with 
different number of atoms:
$
\hat{\rho} = \sum_{n=0}^\infty c_n | \psi_n \rangle \langle \psi_n |
$
and the extension of Eq.\ \ref{eq:hierarchy_small} is straightforward:
\begin{eqnarray}
& &
	\int d\xx_{r} \; G^{(r)}_{\hat{\rho}} (\xx_1,\ldots,\xx_r) = 
\nonumber \\ & &
	\sum_{n=0}^\infty c_n \left( n - r + 1 \right) 
	G^{(r-1)}_{|\psi_n\rangle} (\xx_1,\ldots,\xx_{r-1}) 
	\label{eq:hierarchy_small_mixed}
\end{eqnarray}
Other states, allowing for coherence between states with different number of atoms, violate
the superselection law that stems from the baryonic charge conservation \cite{Weinberg} and thus are unphysical 
for atomic systems.

Joint probabilities of detecting atoms at $\xx_1, \xx_2, \ldots, \xx_r$ form hierarchy analogous to Eq.
\ref{eq:hierarchy_small} and therefore:
\begin{equation}
p^{(r)}_{|\psi\rangle} (\xx_1,\ldots,\xx_r) = 
\frac{({\cal{N}}-r)!}{{\cal{N}}!} G^{(r)}_{|\psi\rangle} (\xx_1,\ldots,\xx_r)
\end{equation}

Let us now investigate the experiment in greater detail and concentrate on an ideal case first, 
i.e.\ when the detectors have infinite spatial resolution 
	and are able to detect individual atoms. 
We will also assume that the detectors' mesh is so tight that each and every atom gets detected. 
Then, photographing the state $|\psi\rangle$ produces a collection of ${\cal{N}}$ positions 
$\xi_1^i, \xi_2^i, \ldots, \xi_{\cal{N}}^i$ at which atoms were found at the $i$-th shot. 
Repeating the measurement ${\cal{W}}$ times follows the corresponding probability density and therefore
yields a discrete estimate of the ${\cal{N}}$-th correlation function:
\begin{equation}
{\cal{G}}^{({\cal{N}},{\cal{W}})}_{|\psi\rangle} (\xx_1,\ldots,\xx_{\cal{N}})
	= \frac{1}{\cal{W}} \sum_{i=1}^{\cal{W}} {\cal{N}}! \: \prod_{j=1}^{\cal{N}} \delta \! \left(\xx_j-\xi_j^i\right)
\end{equation}
Of course in the limit of repeating the experiment infinitely many times the original correlation
function is recovered: 
\begin{equation}
{\cal{G}}^{({\cal{N}},{\cal{W}})}_{|\psi\rangle} (\xx_1,\ldots,\xx_{\cal{N}})
\stackrel{{\cal{W}} \rightarrow \infty}{\longrightarrow}
G^{({\cal{N}})}_{|\psi\rangle} (\xx_1,\ldots,\xx_{\cal{N}})  
\end{equation}
To obtain the density out of these measurements one needs only to integrate the constructed correlation
function over ${\cal{N}}-1$ coordinates according to the Eq.\ \ref{eq:hierarchy_small}. 
(It is assumed that this equation holds true  also for the discrete estimates of correlation functions.)
Since it doesn't matter over which ${\cal{N}}-1$ variables the integration is performed
one obtains a very intuitive result:
\begin{equation}
{\cal{G}}^{(1,{\cal{W}})}_{|\psi\rangle} (\xx) = \frac{1}{\cal{W}} \sum_{i=1}^{\cal{W}}
\sum_{j=1}^{\cal{N}} \delta \! \left(\xx-\xi_j^i\right)
\end{equation}
The density is thus just a sum of $\delta$-spikes at positions at which atoms were detected in all experiments. 
Note the self averaging phenomenon: typically the number of atoms ${\cal{N}}$ is of the order of 
hundreds of thousands and one does not need to repeat the experiments very many
times to obtain a reasonable density approximation. However, 
in general it is not sufficient to perform the experiment only once, no matter how many atoms there are.
The "density" obtained in a {\bf single} measurement, ${\cal{G}}^{(1,1)}_{|\psi\rangle} (\xx)$, is --apart
from the systems exhibiting coherence up to the ${\cal{N}}$-th order--
different from a true density $G^{(1)}_{|\psi\rangle} (\xx) = {\cal{G}}^{(1,\infty)}_{|\psi\rangle} (\xx)$, 
since the first one also carries information about correlations in the system \cite{Gajda}. 

Second order correlation function can be extracted from the measurements' outcome in an analogous way
via integrating
${\cal{G}}^{({\cal{N}},{\cal{W}})}_{|\psi\rangle} (\xx_1,\ldots,\xx_{\cal{N}})$
over ${\cal{N}}-2$ variables. After exploiting the symmetry trick, just as before, we obtain:
\begin{eqnarray}
& &
{\cal{G}}^{(2,{\cal{W}})}_{|\psi\rangle} (\xx,\xx') = \frac{1}{\cal{W}} \sum_{i=1}^{\cal{W}}
\nonumber \\ & &
\sum_{j=1}^{\cal{N}} \sum_{j'=1}^{\cal{N}} \delta \! \left(\xx-\xi_j^i\right) 
\delta \! \left(\xx'-\xi_{j'}^i\right) \left( 1 -\delta_{j,j'} \right)
\end{eqnarray}
This idea has been already used in \cite{Lukin}, in which the authors propose to look at spatial noise
correlations in order to extract second order correlation function.
Similar method can be applied to extract even higher order correlation functions.

To describe realistic experiments --- with inevitably imperfect detectors ---
we will incorporate the following strategy:
For each outcome of a realistic measurement, the set of all possible  ideal realizations (discussed above) 
that are identical from the point of view of 
imperfect detectors, is found; then, for each of these realizations  
the correlation function of interest is calculated
and finally averaging over these realizations takes place.

The outcome of the $i$-th experiment is modelled by a set of ${\cal{D}}$
numbers denoting how many atoms were found by each of the detectors:
$\Xi_1^i, \Xi_2^i, \ldots, \Xi_{\cal{D}}^i$. Here,
the detectors are assumed to have a finite spatial resolution (each of them covering a volume $V$) 
and to fill all available physical space.
Furthermore, they are characterized by the conditional probability
$p(\eta|\Xi)$ of the presence of $\eta$ atoms while $\Xi$ were actually detected. Denoting:
$
 \overline{\eta^k(\Xi)} := \sum_{\eta} p(\eta|\Xi) \eta^k
$
(for example, if each atom is independently detected with a probability $p$ then 
 $\overline{\eta(\Xi)} = \frac{\Xi + 1-p}{p}$ and $\overline{\eta^2(\Xi)}=\overline{\eta(\Xi)}^2 +
 \frac{(\Xi+1)(1-p)}{p^2}$), 
first- and second-order correlation functions take the form:
\begin{equation}
{\cal{G}}^{(1,{\cal{W}})} (\xx) = \frac{1}{\cal{W}} \sum_{i=1}^{\cal{W}}
\sum_{j=1}^{\cal{D}} \overline{\eta(\Xi_j^i)} \Delta_j(\xx),
\end{equation}
\begin{eqnarray}
& & {\cal{G}}^{(2,{\cal{W}})} (\xx,\xx') = \frac{1}{\cal{W}} \sum_{i=1}^{\cal{W}}
\sum_{j=1}^{\cal{D}} \Delta_j(\xx) 
\\ & &
	\left[ \Delta_{j}(\xx') 
\left( \overline{\eta^2(\Xi_j^i)} - \overline{\eta(\Xi_j^i)} \right) + 
\overline{\eta(\Xi_j^i)} \sum_{j'\neq j}^{\cal{D}} \Delta_{j'}(\xx') \overline{\eta(\Xi_{j'}^i)}
\right]
\nonumber
\end{eqnarray}
Here, $\Delta_j(x) = V^{-1}$ if $x$ is within the $j$-th detector and zero otherwise. 
Note that the knowledge of the underlying quantum state is not required here and
still lower order correlation functions can be extracted.

One more remark about experimental reality. A typical measurement performed with cold atoms involves
so called column averaging, in which information about correlations in the direction along the
propagation of the illuminating light pulse is lost. The scheme of extracting correlation functions
presented above is not sensitive to whether the original high-order correlation function is column averaged or
not, and in this sense this process is not an obstacle. But if one starts from a column averaged high-order
correlation function one ends up with column averaged low-order correlation functions. In many situations
this is not satisfactory, however, because
such an averaging kills interesting phenomena. Consider, for example, the recent experiment of collapses and
revivals of matter wave field of a Bose-Einstein condensate \cite{Immanuel2}. Only because of the column averaging
were the collapses of the interference pattern observed in a single shot. If one repeated the experiment
with 2-dimensional lattices, interference patterns would be observed at each measuerement no matter what the
free evolution time was and they would vanish only when averaged over many realizations. 

Let us illustrate this general method of extracting higher order correlation functions
with an example of a coherent superposition of states composed of ${\cal{N}}$ atoms of the form:
\begin{equation}
\left| \psi \right\rangle_{COH} = \frac{1}{\sqrt{2}} \left( \left|{\cal{N}},0\right\rangle + 
		\left|0,{\cal{N}}\right\rangle \right)
\label{eq:cat}
\end{equation}
Such a state provides a very important link between quantum and classical worlds, is essential 
in the theory of quantum information, and is currently being a subject of intensive research \cite{Raizen}. 
We will assume that the wavefunctions associated with the two modes are the ground states of spatially
separated potential wells: $\varphi_L(x)$ and $\varphi_R(x)$. The $r$-th order correlation function is: 
(a) if $r<{\cal{N}}$:
\begin{eqnarray}
& & G^{(r)}_{COH} (x_1,x_2,\ldots,x_r) = 
\nonumber \\ & &
\frac{{\cal{N}}!}{2 ({\cal{N}}-r)!} \left( 
\prod_{i=1}^r \left| \varphi_L(x_i) \right|^2 + \prod_{i=1}^r \left| \varphi_R(x_i) \right|^2
\right),
\label{eq:rcorrfunCOH}
\end{eqnarray}
(b) if $r={\cal{N}}$:
\begin{equation}
G^{({\cal{N}})}_{COH} (x_1,\ldots,x_{\cal{N}}) = 
\frac{{\cal{N}}!}{2} \left| \prod_{i=1}^{\cal{N}} \varphi_L(x_i) +
\prod_{i=1}^{\cal{N}} \varphi_R(x_i) \right|^2
\label{eq:NcorrfunCOH}
\end{equation}
and (c) if $r>{\cal{N}}$ the correlation functions vahish indentically.

Explicit formulas for correlation functions allow us to simulate the experiment easily.
Since each measurement produces a set of numbers drawn according to the corresponding joint
probability function, so can it be done in the computer with the help of random number generator. 
This is achieved step by step: the first atom is drawn with the
probability density given by $G^{(1)}_{COH}(x_1)$, the second -- with the probability density
$G^{(2)}_{COH}(x_1=\xi_1,x_2)$, where for $x_1$ a number obtained in the previous step is substituted 
and so on \cite{Javanainen}. In this way the full set $\xi_1,\xi_2,\ldots,\xi_{\cal{N}}$ is obtained. 

\begin{figure}
\epsfig{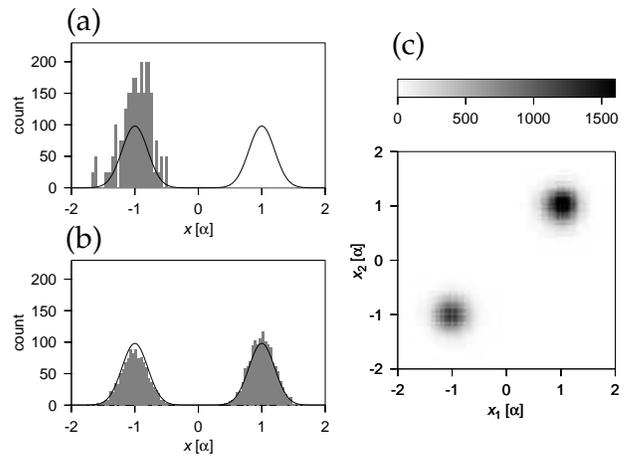}
\caption{Low-order correlation functions for a superposition state $\left| \psi \right\rangle_{COH}$
	with ${\cal{N}}=100$ atoms:
	(a) density obtained in a single measurement, (b) density averaged over 50 shots and (c) second order correlation
		function averaged over 50 measurements.
The wavefunctions $\varphi_{L,R}(x)$ were initially Gaussian wavepackets centered at $x_0=\pm \alpha$ 
($\alpha$ denotes the unit of length) with width $\sigma=0.035 \alpha$ to make them practically orthogonal,
	but they were allowed to expand freely for  time $t=0.01 \frac{m \alpha^2 }{\hbar}$ before the detection
process. The solid line at (a) and (b) shows the expected density profile, $G^{(1)}_{COH} (x)$. }
\label{fig:density}
\end{figure}

Figure \ref{fig:density} shows density and second order correlation function 
obtained via the hereinbefore prescription from the numerically 
simulated measurements. Fig.\ \ref{fig:density}(a) 
	shows a typical result of a single measurement, in which all atoms are detected
in one of the wells. Averaging over many experiments leads to the correct density profile, as shown 
at Fig.\ \ref{fig:density}(b).
Fig.\ \ref{fig:density}(c) shows the second order correlation function constructed from 
single measurements and then averaged over many shots.

\begin{figure}
\epsfig{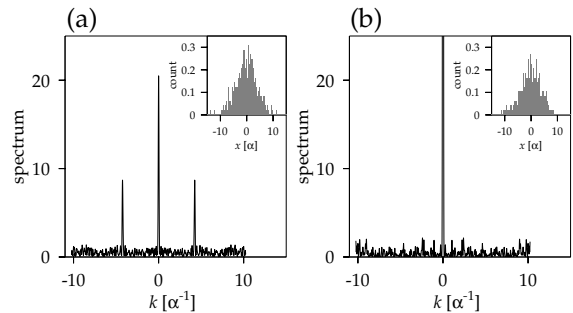}
\caption{Absolute value of Fourier transforms of histograms of the center-of-mass variable for 
	(a) coherent superposition and (b) mixed state for  ${\cal{N}}=20$ atoms and 1 000 measurements
		(histograms are shown in the insets).  
		The spikes at $k= \pm \frac{{\cal{N}}}{\pi} \frac{t (m x_0/\hbar)}{t^2 + (m \sigma^2/\hbar)^2}$, which
		for our set of parameters correspond to $k=\pm 4.24$,
	   are fingerprints of the coherence between states $|{\cal{N}},0\rangle$ and $|0,{\cal{N}}\rangle$.}
\label{fig:centerofmass}
\end{figure}

It is instructive to consider another problem: suppose we want to distinguish between the coherent superposition
defined above and the mixed state:
\begin{equation}
\rho = \frac{1}{2} \left(  
		\left| {\cal{N}}, 0 \right\rangle \left\langle {\cal{N}}, 0 \right|
		+ \left| 0, {\cal{N}} \right\rangle \left\langle 0, {\cal{N}} \right|
		\right)
\end{equation}
(Similar problem for four ions in a trap was experimentally investigated in \cite{Sackett}.)
These states have all correlation functions identical apart from the one of
the order of the number of particles: for mixed state the Eq.\ \ref{eq:rcorrfunCOH} holds even when $r={\cal{N}}$.
It follows immediately that in order to distinguish between these states an experiment measuring all atoms 
must be performed in conditions in which the spatial wavefunctions overlap, otherwise the interference
term in Eq. \ref{eq:NcorrfunCOH} will vanish. 
Since the correlation function of interest lives in the $d \times {\cal{N}}$-dimensional space ($d$
		is the physical dimension of the space), it is practically impossible to gather satisfactory statistics
of experiments to probe this space even for moderate number of atoms. 
Fortunately, there is a way around it. Our purpose is to differentiate between two kinds of states and any
quantity that does so is enough. According to what has been said, such a quantity must originate from
the ${\cal{N}}$-th order correlation function, but it does not need to be the correlation function itself. 
Simple calculations show that the density of the center-of-mass is indeed such a quantity and a direct proof
obtained from numerical simulations is shown in Figure \ref{fig:centerofmass}. 
Note that all atoms must necessarily be detected to obtain a center-of-mass density, which seems impractical at first. 
However, when performing this kind of experiment a very precise
control of the input state is required, in particular the knowledge of the total number of atoms, and thus
it is natural to discard all measurements that detected less than ${\cal{N}}$ atoms
and this way to overcome finite detector efficiency.

This method of diagnosing the superposition has of course its limitations. When increasing the number of atoms, the
visibility of fringes in the center-of-mass density decreases, as well as the separation between fringes and 
consequently better and better detectors are required. Moreover, the sensitivity to the phase
fluctuations from shot to shot increases: if we assume that the initial state is  proportional to
$ |{\cal{N}},0 \rangle + \exp(i\phi) |0,{\cal{N}}\rangle $, where $\phi$ varies from experiment to
experiment, it is required for the method to work that the phase $\phi$ does not change more than over
$\pi/(2{\cal{N}})$. The number of experiments needed to distinguish between a coherent superposition
and a mixed state grows rapidly with the number of atoms, 
	but still yields a reasonable number of repetitions: under the condition that the peaks' height signifying
	the coherent superposition is 10 times the value of surrounding background, $500$ runs are required 
	for ${\cal{N}}=20$, while $5 000$ for ${\cal{N}}=100$ -- and bigger coherent superpositions are not only 
difficult to diagnose, but even more, to prepare.

Of course the mesoscopic superposition prepared even in the best experiments cannot be expected to be
exactly of the form given by Eq. \ref{eq:cat} but rather of the form $\sum_{k=0}^{\cal{N}} c_k |k,{\cal{N}}-k \rangle$,
where the coefficients $c_k$ are peaked around $k=0$ and $k={\cal{N}}$. The state we have discussed above
has been chosen for its simplicity, but the analysis may easily be extended to more
realistic states.

Summarizing: the fact that atomic systems cannot exist in a superposition of Fock states with
different number of atoms leads to a very simple hierarchy of correlation functions. Basing on that we have 
provided a prescription how to extract higher order correlation functions from typical measurements performed in
atom optics. We have 
discussed both the idealized and realistic version of experiments and illustrated the method by applying it to
a superposition of atomic states. We have also proposed a method of diagnosing such a superposition 
(versus an incoherent
mixture) via measuring the center-of-mass density profile.

\begin{acknowledgements}
We would like to thank K. Banaszek, M. Gajda and J. Mostowski for their useful comments.
The work was partially supported by the Subsidy of Foundation for Polish Science and by the Polish Ministry of
Scientific Research and Information Technology under grant No PBZ-MIN-008/P03/2003.
\end{acknowledgements}

\end{document}